\documentclass[aps,pre,showpacs,groupedaddress,twocolumn,showkeys,amsmath,amssymb]{revtex4}

\usepackage{graphicx}
\usepackage{epsfig}
\usepackage{bm}
\usepackage{bbm}

\usepackage{epic}
\usepackage{eepic}
\usepackage{pifont}
\usepackage[usenames]{color}

\usepackage{nicefrac}
\def\vec#1{\mathbf{#1}}

\input epsf
\input rotate
\usepackage{graphicx}

\begin{document}
\draft

\title{Vesicles in a Poiseuille flow}
\author{Gerrit Danker$^1$, Petia M. Vlahovska$^2$, and Chaouqi Misbah$^1$}
\affiliation{$^1$Laboratoire de Spectrom\'etrie Physique, UMR, 140
avenue de la physique, Universit\'e Joseph Fourier, and CNRS,  38402
Saint Martin d'Heres, France\\
$^2$ Thayer School of Engineering, Dartmouth College, 8000 Cummings
Hall, Hanover NH 03755, USA}

\email[]{chaouqi.misbah@ujf-grenoble.fr}
\date{\today}
\begin{abstract}

Vesicle dynamics in unbounded  Poiseuille flow is analyzed using a small-deformation theory. Our analytical
results quantitatively describe vesicle migration and provide new physical insights.
At low ratio between the inner and outer viscosity
$\lambda$ (i.e. in the tank-treading regime),
the vesicle always migrates towards the flow centerline, unlike other soft particles such as drops.
Above a critical $\lambda$, vesicle  tumbles and cross-stream migration vanishes.
A novel
feature is predicted, namely
   the coexistence of two types of nonequilibrium configurations  at the centreline, a bullet-like
  and a parachute-like shapes.

\end{abstract}


\maketitle

The motions of vesicles made of closed bilayer membranes in viscous flows have received increasing attention in recent years
because of their relevance to understanding cell dynamics in the microcirculation. The unusual mechanical properties of the lipid bilayer membrane,  such as fluidity, incompressibility and resistance to bending,
give rise to a number of fascinating nonequilibrium features of vesicle micro-hydrodynamics
For example, in linear shear flow vesicles are predicted to exhibit
different types of motion: (i) tank-treading (TT) (the fluid
membrane rotates as a tank-tread, while the orientation angle of the
vesicle remains fixed in time) \cite{Kraus1996,Seifert1999}, (ii)
tumbling (TB) \cite{Keller1982,Biben2003}, (iii)
vacillating-breathing (VB) (the long axis undergoes oscillation
about the flow, while the shape shows breathing) \cite{Misbah2006}.
Some of these behaviors have been confirmed experimentally
\cite{deHaas1997, Kantsler2005, Mader2006}. Vesicle dynamics in
linear flows is now fairly well understood
\cite{Seifert1999,Misbah2006,Vlahovska2007,Lebedev2007,Gompper2007,Danker2007b}.

Quadratic flows such as the Poiseuille flow are quite common,
especially in the blood circulatory system, yet the impact of flow curvature on vesicle dynamics is not well understood.
Number of studies have focused on capillary flows, where the vesicle diameter is comparable to the
channel size and lubrication effects control vesicle motion \cite{Vitkova, Noguchi-Gompper:2005b},
 or vesicles near walls, where hydrodynamic interactions with the wall
  give rise to cross-stream migration (in a direction perpendicular to the flow)
 \cite{Olla, Abkarian:2005}.
 Blood vessels such as arterioles, however, can be 10-50 times larger than the cell diameter. In this case,
  effects of flow curvature may become dominant. Vesicle dynamics in unbounded Poiseuille flow has been
  considered only to a limited extent, using two-dimensional numerical simulations \cite{Badr}.
 Cross--stream migration has been observed,
   even in the absence of wall \cite{Badr}. The physical mechanism behind this migration is
   related to the spatial variations of shear rate that exists in quadratic flows:  deformable particles tend
   to move towards regions with lower shear in order to minimize
    shape distortion \cite{Leal1980}.
 Drops, for example, migrate
 towards or away the flow centerline depending on the viscosity ratio \cite{Leal1980}. In the case of vesicles, however,  details
 of the migration mechanism remain elusive and  there are
 number of open questions: What is the role of flow curvature in cross-stream vesicle migration?
 Is there relation between the migration direction and the basic modes of vesicle dynamics  (i.e. TT,
VB and TB)? What physical parameters control vesicle migration? The
development of a theory that can answer these questions represents a
challenging task because not only the shape, but also the location
of the vesicle is not known a priori, and it must be solved for in a
consistent manner.

{\it{ Problem formulation and solution:}}
In a reference frame centered in the vesicle, a channel flow is written as
\begin{eqnarray}
  \vec v_0 = ( -v_\mathrm{s}-\dot\gamma y -\alpha y^2)\,\vec e_x ,
  \label{eq:Pois2}
\end{eqnarray}
 where  $\alpha$ is
a measure of the curvature of the flow profile, $\dot\gamma$ is the
local shear rate, which depends to the distance between the vesicle
center and the flow axis,  $\dot \gamma=2 y_0 \alpha$,  and
$v_\mathrm{s}$ is the slip velocity (to be determined), which is the
difference between the actual velocity of the vesicle center in the
flow direction and the unperturbed flow.

At the length scales of the vesicle, water is effectively very viscous. Hence,  the total flow field $\vec v = \vec v_0 + \vec u$ (imposed
flow $\vec v_0$ plus perturbation $\vec u$ due to the presence of
the vesicle) obeys the Stokes equations inside and outside the vesicle
\begin{equation}
  \nabla p = \eta_i\nabla^2 \vec v,\quad
  \nabla\cdot\vec v = 0,
  \label{eq:floweqs}
\end{equation}
where $\eta_i$ ($i=1,2$) is the fluid viscosity;
$\eta_1,\eta_2$ are the internal and external viscosities, and
$\lambda\equiv \eta_1/\eta_2$ is a measure of the viscosity
contrast.

Assuming a nearly spherical shape, the solution to the
creeping--flow equations is obtained as a regular perturbation
expansion in the excess area, which is the difference in the areas
of the vesicle and an equivalent sphere (with radius $r_0$) with the
same volume, i.e., $\Delta=A/r_0^2-4\pi$.
From the definition of the area it is evident that $\Delta$ is a quadratic function of the shape deviation from sphere.
Thus it is convenient to set $\epsilon=\Delta^{1/2}$  as a formal expansion parameter.
A time-dependent vesicle configuration is described by
\begin{eqnarray}
  \vec R = \vec R_0 + r_0 [1 + \epsilon f(t,\theta,\phi)]\,\vec
  e_r,
  \label{eq:R}
\end{eqnarray}
where $\vec R_0 = (x_0, y_0, z_0)$ is the time-dependent position of
the centroid  of the vesicle and $f$ describes shape deviation from
the spherical shape. $\theta$ and $\phi$ are spherical coordinates.
The vesicle shape $f(\theta,\phi)$ can be decomposed in an infinite
series of spherical harmonics $Y_{lm}(\theta,\phi)$ excluding the
$l=1$ mode (which describes the translation $\dot{\vec R}_0$):
\begin{eqnarray}
  f = \sum_{l\neq1} f_l = \sum_{l\neq1} \sum_{m=-l}^l
  F_{lm}(t)\,Y_{lm}(\theta,\phi).
  \label{eq:f}
\end{eqnarray}
The amplitudes $F_{lm}(t)$ describe the shape evolution.
For the present study (leading order analysis) it is sufficient to include
the $l=0$ mode, which ensures volume conservation, and the $l=2,3$
modes, which are excited by the imposed Poiseuille flow.

The  velocity field is given by the classical Lamb
solution \cite{Lamb1932}. This solution contains integration factors
which are determined from the boundary conditions. These are:  (i)
Continuity of the fluid velocity across the membrane (valid if
impermeability is assumed).  (ii) Continuity of the stress across
the membrane: the jump of fluid stress across the membrane is
balanced by the membrane force. The latter consists of  a normal
force due to resistance to bending, and a tangential (fictitious)
force, which originates from a Lagrange multiplier enforcing local
membrane incompressibility. (iii) Local membrane incompressibility,
which restricts the surface velocity field to be solenoidal.
We have solved the full hydrodynamic problem in the same spirit
as
 in the calculation for the linear shear
flow. Details can be found in \cite{Danker2007b}; here we
focus solely on the  results and discuss the physical implications.

It should be noted that in certain cases, e.g., drops,  the migration velocity can be
obtained using a simplified approach based on the Lorentz reciprocal
theorem\cite{ChanLeal1979}. However, this approach is inapplicable
to vesicles because of the tangential membrane force.

{\it{Results and Discussion:}}
The time evolution of the shape function
($\partial_t f$) together with the motion of the vesicle center with
respect to the flow ($\dot{\vec R}_0$) are computed from the kinematic condition, which requires that the
membrane moves with the  fluid velocity.
We find for the evolution of the $l=2,3$ (which are the only
excited modes at leading order) the following equations:
\begin{eqnarray}
  \label{eq:dtf2}
  \epsilon\,{\cal D}_t F_{2m} &=& -
  \frac{24\,(\sigma_0 + 6\kappa)}{23\lambda+32} F_{2m} +
  \frac{\mathrm{i} \alpha y_0}{23\lambda+32} N_{2m},
  \\
  \label{eq:dtf3}
  \epsilon\,{\cal D}_t F_{3m} &=&
  -\frac{120\,(\sigma_0 + 12\kappa)}{76\lambda+85} F_{3m} +
  \frac{\alpha r_0}{76\lambda+85} N_{3m}
\end{eqnarray}
with  ${\cal D}_t = \partial_t + \mathrm{i} m\alpha y_0$, and
$N_{20} = N_{2,\pm 1} = 0$, $N_{2,\pm 2} = \pm 8\sqrt{30\pi}$,
$N_{30} = N_{3, \pm 2} = 0$, $N_{3,\pm 1} = \mp 5\sqrt{21\pi}/3$,
$N_{3,\pm 3} = \mp 5\sqrt{35\pi}$.
$\kappa$ is the membrane bending rigidity rescaled by $r_0^3/\eta$.
The evolution equations (\ref{eq:dtf2}, \ref{eq:dtf3}) contain the
tension-like quantity $\sigma_0$, which is the homogeneous part of
the Lagrange multiplier \cite{Seifert1999,Danker2007b}. Its value is
computed from the condition that the shape deformation complies with
the available excess area. $\sigma_0$ is easily determined from the
above equations, and the expression relating excess area and the
shape modes $F_{lm}$, as explained in \cite{Danker2007b}.

Knowledge of the shape evolution allows us to compute  the lateral (cross-stream) migration
along the $y$-direction
\begin{eqnarray}
  v_\mathrm{m} &=& \frac{\mathrm{i}}{4} \sqrt{\frac{30}{\pi}} \alpha\epsilon\,
  r_0^2 \frac{36\lambda+71}{(\lambda+4)(76\lambda+85)} (F_{22} -
  F_{2,-2}) \nonumber\\
  && {}+ \frac{16}{7}\sqrt{\frac{21}{\pi}} \frac{\alpha\epsilon\, r_0
  y_0}{23\lambda+32} (F_{31} - F_{3,-1}) \nonumber\\
  && {}+ \frac{48}{7}\sqrt{\frac{35}{\pi}} \frac{\alpha\epsilon\, r_0
  y_0}{23\lambda+32} (F_{33} - F_{3,-3}).
  \label{eq:vy}
\end{eqnarray}
This equation is coupled back to the shape, leading to
some interesting dynamics. First, in contrast to drops,
a vesicle always migrates towards the centerline. Second, the migration velocity depends in a non-trivial way
on the viscosity ratio and excess area.

\begin{figure}[ht]
  \begin{center}
    \epsfig{figure=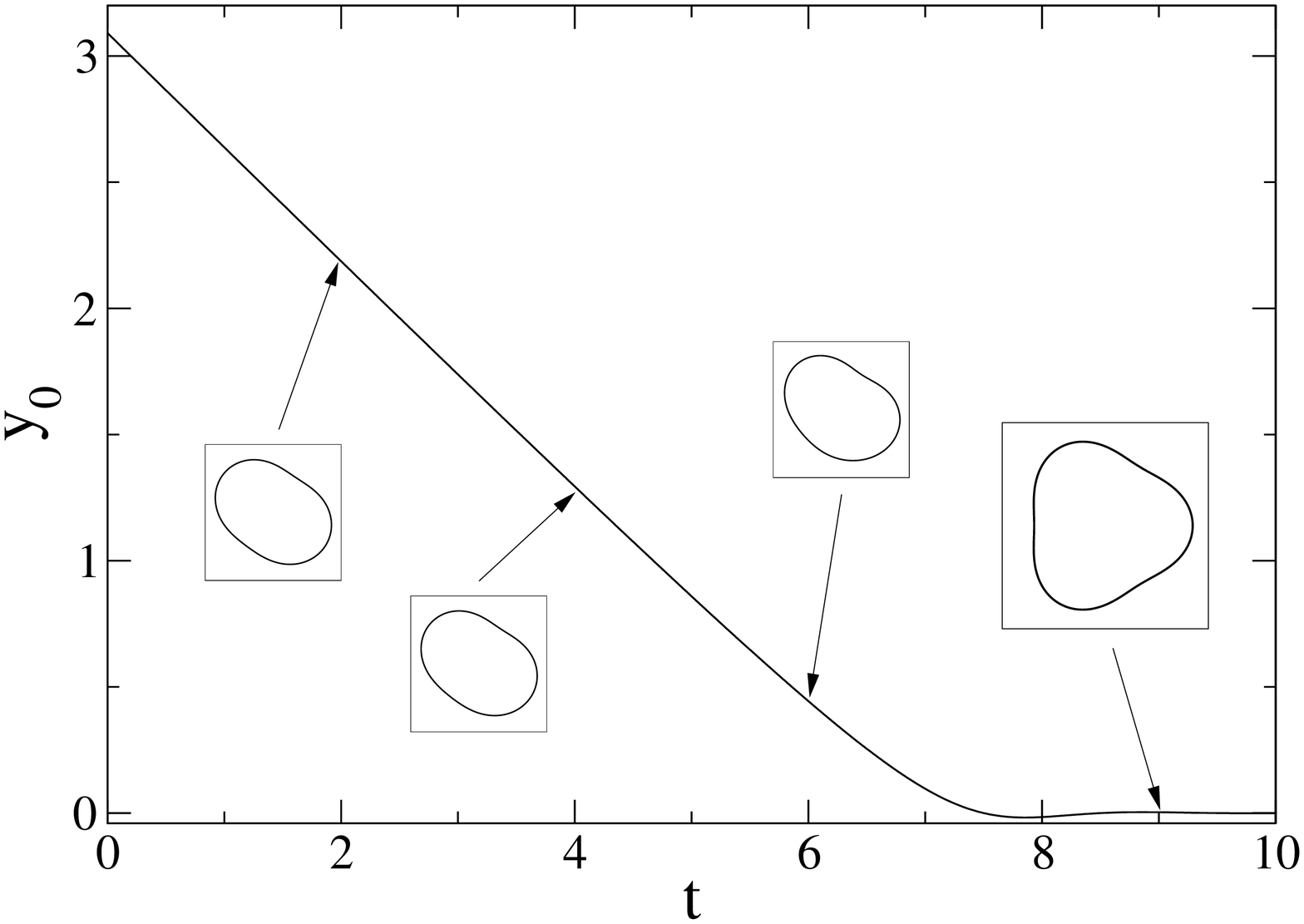,width=.70\linewidth}
  \end{center}
  \caption{Evolution of the vertical position of the vesicle as a function of
  time.  Some snapshots of the vesicle shape along the trajectory are
  shown as insets.  Parameters:  $\alpha=6, \kappa=1, \Delta=0.25$.}
  \label{fig:y0shapes}
\end{figure}

For $\lambda < \lambda_\mathrm{c}$, a
numerical solution of the coupled equations for the shape
(\ref{eq:dtf2}, \ref{eq:dtf3}) and the migration (\ref{eq:vy})
yields the usual tank-treading ellipsoid as long as the vesicle is
far from the centerline, because the linear shear dominates over the quadratic component of the flow.  The latter, however,
 induces migration towards the center of
the flow.  As the vesicle moves its
shape deforms more and more until it becomes a parachute (or assumes
another shape, namely bullet-like, as discussed below)  when the
centerline is reached (Fig.~\ref{fig:y0shapes}).
The migration velocity far from the centerline is well approximated by plugging
the solution for  the tank-treading ellipsoidal shape, $R=\epsilon/2$, $\cos 2\psi = \epsilon/2h$ with $R$ and $\psi$
defined by $\epsilon F_{22} = R\,e^{-\mathrm{i}2\psi}$) into
Eq.~(\ref{eq:vy})
\cite{Misbah2006}.
\begin{eqnarray}
  v_\mathrm{m} =
  -\frac{17284\lambda^2 + 66671\lambda +
  55840}{60\,(\lambda+4)(23\lambda+32)(76\lambda+85)}
  \sqrt{\frac{30}{\pi}}\,\alpha\Delta^{1/2}\,r_0^2.
  \label{eq:vy_scal2}
\end{eqnarray}
For $\lambda\sim 1$ the migration velocity is approximately $ v_y =-
0.16\,\alpha\Delta^{1/2} r_0^2.  $ In a typical experiment in a
rectangular microchannel we have $\alpha\sim
0.3\,(\mbox{$\mu$m$\cdot$s})^{-1}$, $\Delta = 0.25$, $r_0 =
20\,\mbox{$\mu$m}$, and a migration velocity of
$0.75\,\mbox{$\mu$m/s}$\cite{Coupier07}.  Using
Eq.~(\ref{eq:vy_scal2}), we find an expected migration velocity of
about $1\,\mbox{$\mu$m/s}$. Note also that (\ref{eq:vy_scal2}) is
independent of the membrane bending rigidity. This is understood as
follows. The local shear rate is proportional to the distance from
the centerline:  $\dot\gamma\sim y_0$. The elongational component of
the flow (which increases like the shear rate) must be compensated
by the membrane tension in order to fulfil local membrane
incompressibility.  It follows that the tension $\sigma_0$ must also
scale like $y_0$.  Then, for large $y_0$, the terms $6\kappa$ and
$12\kappa$ in Eqs.~(\ref{eq:dtf2}, \ref{eq:dtf3}) become negligible
and the membrane dynamics becomes independent of $\kappa$.

\begin{figure}[ht]
  \begin{center}
    \epsfig{figure=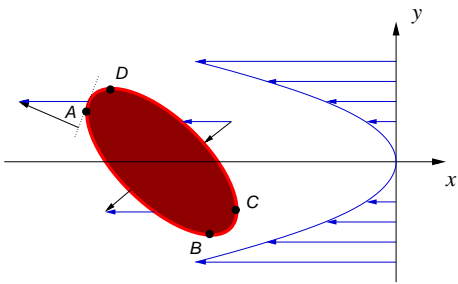,width=.7\linewidth}
  \end{center}
  \caption{(Color online)  A vesicle in Poiseuille flow far from the centerline.  The
  local velocity field can be decomposed into a local shear
  $\dot\gamma(y_0)$ (responsible for the tank-treading vesicle at the
  shown orientation) and a (small) quadratic correction (blue arrows).
  Due to the inclination of the vesicle long axis with respect to the
  quadratic flow field, there is a net force acting on the vesicle
  membrane in the negative $y$ direction.}
  \label{fig:migration}
\end{figure}

The direction of the lateral migration of the vesicle in Poiseuille flow can
intuitively be understood from the following argument.
Assume that we have solved the problem for the pure shear flow,
$\dot\gamma(y_0)y$.  The solution is the tank-treading vesicle shown in
Fig.~(\ref{fig:migration}).
Next, we add the quadratic correction $-\alpha y^2$ as a
perturbation. The additional flow, acting on the membrane of the
vesicle, can be decomposed into two components:  one tangential to
the membrane, which  locally modifies the tank-treading velocity,
and one normal to the membrane, which modifies the vesicle
shape and possibly cause migration.  In the segments $AB$ and $CD$
the perpendicular component tends to push the vesicle into the
negative $y$ direction, whereas in the smaller segments $DA$ and
$BC$, the perpendicular component pushes the vesicle in the positive
$y$ direction.  As the segments $AB$ and $CD$ account for most of
the vesicle's surface, the net force has a component in the negative $y$
direction (towards the flow centerline).

\begin{figure}[t]
  \begin{center}
    \epsfig{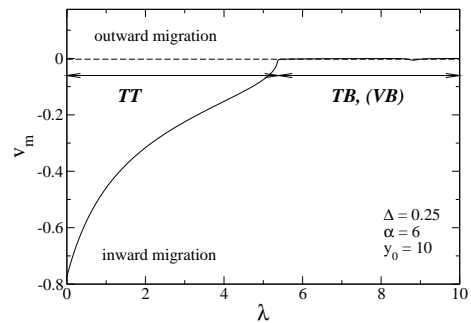}
  \end{center}
  \caption{Migration velocity as a function of the viscosity contrast
  $\lambda$ for the various regimes:  tank-treading (TT), tumbling (TB), and
  vacillating-breathing (VB).}
  \label{fig:vmigration}
\end{figure}

Let us  turn now to the problem of migration in the tumbling regime.
From the analytical  results discussed above for the tank-treading
regime, one finds that the migration velocity close to the tumbling
regime (which occurs at $\lambda=\lambda_c$ \cite{Misbah2006})
behaves as $\sqrt{\lambda-\lambda_c}$, and thus it vanishes exactly
at the threshold. This analysis is valid on the tank-treading side.
For $\lambda>\lambda_c$, the vesicle tumbles, and in this case we
have integrated numerically the set of equations
(\ref{eq:dtf2},\ref{eq:dtf3},\ref{eq:vy}). The
migration (averaged over time) is again always towards the flow centerline but
the magnitude of the migration velocity is very close to zero (of
the order of $10^{-2}\,r_0$ per tumbling cycle). The results are
shown on Fig.\ref{fig:vmigration}.
\begin{figure}[t]
  \begin{center}
    \epsfig{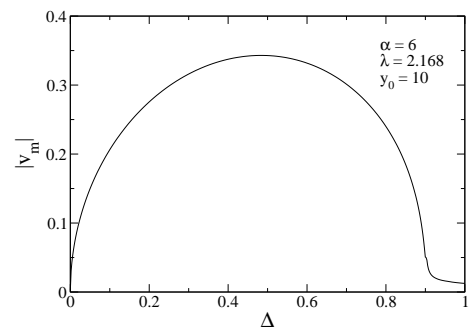}
  \end{center}
  \caption{Migration velocity as a function of the excess area far from
  the centerline.  Parameters are chosen such that the tank-treading regime stops at
  $\Delta = 0.9$.}
  \label{fig:vmdelta}
\end{figure}
\begin{figure}[t]
  \begin{center}
    \epsfig{figure=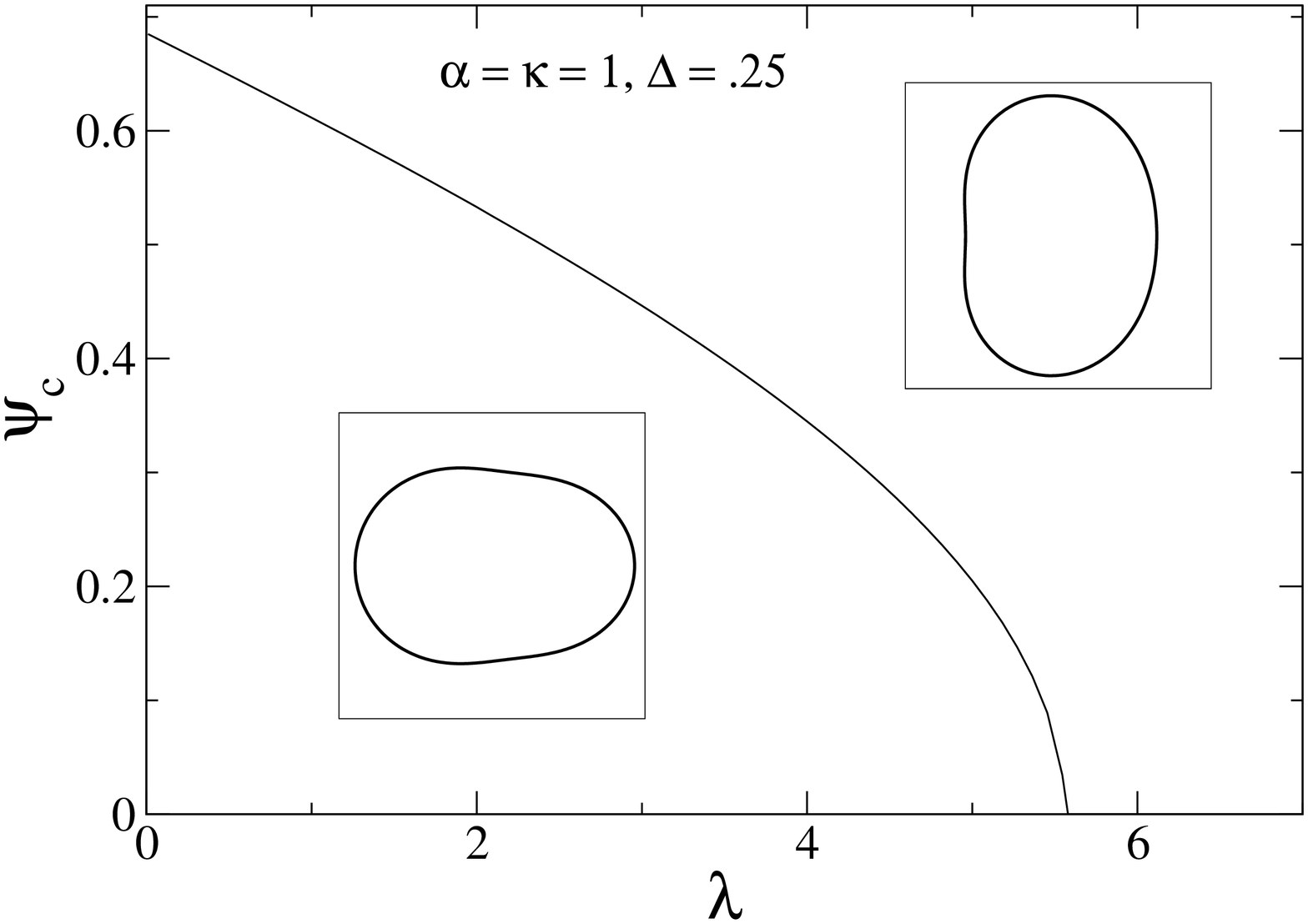,width=.70\linewidth}
  \end{center}
  \caption{Critical angle $\psi_\mathrm{c}$ separating the basins of
  attraction for two coexisting solutions at the centerline.}
  \label{fig:sepangle}
\end{figure}
For large viscosity ratios, $\lambda\gg 1$, the migration velocity must go to
zero, since the vesicle behaves as a rigid sphere.
The {kinematic} reversibility of the Stokes equations
precludes a lateral migration in this case.
 This feature is  also {evident from} the expression for the migration velocity
(\ref{eq:vy}). Indeed, the $F_{ij}'s$ are finite
due to the constraint of fixed  excess area, but the prefactors $O(\lambda^{-1})$. Thus, when
$\lambda\rightarrow \infty$, $v_\mathrm{m}\rightarrow 0$.

An interesting feature is { the non-monotonic dependence} of
the migration velocity
on the excess area $\Delta$, see Fig. ~\ref{fig:vmdelta}.
For a spherical object (i.e.,
$\Delta=0$) one has  no migration, $v_\mathrm{m} = 0$, due to the up-down symmetry.
If viscosity is large enough so that tumbling becomes possible for a
sufficiently deflated vesicle, then we similarly expect
$v_\mathrm{m} \sim 0$, for high enough $\Delta$.
Hence, it follows that the absolute value of the migration velocity attains
a maximum for a certain value of $\Delta$ as illustrated in
Fig.~\ref{fig:vmdelta}. This results also agrees with the argument
given on the lift force under a linear shear flow\cite{Olla}. Vesicle approach to the flow centerline also can
proceed in a non-monotonous manner.
the ratio $\eta_0 R_0^4\,\alpha/\kappa$ (this is a measure of the
{relative strength of hydrodynamic  and bending stresses}) is
sufficiently large (typically above 10), the vesicle approaches the
centerline monotonously. However, for smaller curvature of the flow
field (or higher membrane rigidity) {the vesicle trajectory exhibits
damped oscillations about the centerline}. For typical values
$\alpha\sim 0.1 (\mu m s)^{-1}$, $R_0\sim 10 \mu m$, and by using
the viscosity of water and $\kappa\sim 40 k_BT$, one finds that
$\eta_0 R_0^4\,\alpha/\kappa \sim 0.1$. This means
these damped oscillations should be  easily
observed experimentally.

{A new surprising feature we have discovered is the coexistence of shape solutions at
the centerline  for small enough curvature of the flow field}  (see
Fig.~\ref{fig:sepangle}).  More precisely,  this occurs if $\eta_0
R_0^4\,\alpha/\kappa$ is lower than about $4-5$. For one solution
the  longest axis of the vesicle is oriented in the flow direction
(bullet-like shape), and for the other one it is perpendicular to it
(parachute-like shape). Each of these solutions has its own basin of
attraction, i.\,e., if the initial angle at the centerline is
smaller than $\psi_\mathrm{c}$, the bullet-like shape is attained,
and otherwise the parachute-like shape. If, however, the curvature
of the flow is strong enough, the vesicle will assume a pronounced
parachute-like shape, and the two solutions will be
indistinguishable. We hope to report further on this matter in a
future work. The experimental range that is relatively easily
accessible is about\cite{Coupier07} $10^{-1}<\eta_0
R_0^4\,\alpha/\kappa<10$. This means that in principle this
prediction is not devoid of experimental testability. The challenge
is to
orient the vesicle in the appropriate basin of attraction. Optical
tweezers constitute  a possible tool for this task.
The slip velocity  for the  parachute-like shape is higher, while for a bullet-like vesicle is
lower than that of a rigid sphere, $v_\mathrm{s} =
-1/3\,\alpha r_0^2$ (as deduced from Faxen's law \cite{slip}).
Since the slip velocity for the bullet-shape is smaller than
that of the parachute the dissipation implied by this morphology
should be lower.

In summary,
the dynamical  behavior of a vesicle in unbounded quadratic flow has revealed number of novel features
 that stem from the subtle interplay between membrane mechanics and shear gradients due to flow curvature. The analysis
has been also performed  for an axisymmetric Poiseuille flow and
yields the same qualitative results.

C.M. Acknowledges financial support from CNES (Centre National
d'Etudes Spatiales) and ANR (MOSICOB project).


\end{document}